\documentclass[pra,aps,amsfonts,amsmath,superscriptaddress,%
  onecolumn,showpacs,floatfix]{revtex4}
\usepackage{amsfonts}
\usepackage{amsmath}
\usepackage{bm}
\usepackage{epsfig}

\begin{document}

\title{\bf The role of superfluidity in nuclear incompressibilities}.

\author{E. Khan}
\address{
{\it Institut de Physique Nucl\'eaire, Universit\'e Paris-Sud,
IN2P3-CNRS, F-91406 Orsay Cedex, France}
}


\begin{abstract}
Nuclei are propitious tools to investigate the role of the superfluidity in
the compressibility of a Fermionic system. The centroid of the Giant
Monopole Resonance (GMR) in Tin isotopes is predicted using a constrained
Hartree-Fock Bogoliubov approach, ensuring a full self-consistent treatment.
Superfluidity is found to favour the compressibitily of nuclei. Pairing
correlations explain why doubly magic nuclei such as $^{208}$Pb are stiffer
compared to open-shell nuclei. Fully self-consistent predictions of the GMR
on an isotopic chain should be the way to microscopically extract both
the incompressibility and the density dependence of a given energy functional.
The macroscopic extraction of K$_{sym}$, the asymmetry incompressibility, is
questioned. Investigations of the GMR in unstable nuclei are called for.
Pairing gap dependence of the nuclear matter incompressibility should also
be investigated. 
\end{abstract}

\maketitle

The effect of the superfluidity on the compressibility of a Fermionic system
remains an open question. Superfluidity initially referred to a system with
a dramatic drop of its viscosity \cite{leg99}: it could be suspected that a
super-fluid would be easier to compress than a normal fluid. This question
has been investigated in Fermionic atoms traps, by studying the frequency of
the compression mode with respect to the scattering length. Experimentally
some increase of the frequency may be observed in the weak pairing regime
\cite{kin04}, but this signal remains to be confirmed. Theoretically both
microscopic and hydrodynamical investigations show no variation of the
compression mode between the normal and the superfluid phases \cite{gra05},
but the analysis is complicated by the temperature change between the two
phases. In nuclear physics, the study of the role of superfluidity in the
compressibility can also be performed: the isoscalar Giant Monopole
Resonance (GMR) is a compression mode, allowing to probe for related
superfluid effects. Ideal tools are especially isotopic chains, where
pairing effects are evolving from normal (doubly-magic) nuclei to superfluid
(open-shell) ones \cite{boh69}. Moreover, the incompressibility of nuclear
matter is a basic parameter in calculations describing neutron stars or
supernovae, where superfluid effects are known to occur \cite{pin85}.

Constraining the nuclear incompressibility modulus K$_\infty$ with
experimental data on the Giant Monopole Resonance is a longstanding problem.
The first relevant approaches have been performed following the work of
Blaizot and Pearson \cite{bla80,pea91}: only microscopic predictions of the
GMR compared with the data could validate the K$_\infty$ value of the
functional which was used. Using a fully self-consistent approach such as
the Random Phase Approximation (RPA) \cite{rs80} or the constrained
Hartree-Fock (CHF) \cite{boh79} method, the GMR data on $^{208}$Pb currently
provides K$_\infty$$\simeq$ 230 MeV in non-relativistic approaches
\cite{col04} whereas K$_\infty$$\simeq$ 260 MeV is obtained for relativistic
one \cite{vre03}. The roots of this puzzle between the two approaches is
still an open question, but it has been shown that the neutron-proton
asymmetry dependence of the incompressibility, denoted as K$_{sym}$, plays
also a role, as well as the density dependence of the functional
\cite{col04b}: a value of K$_\infty$$\simeq$ 250 MeV could be extracted from
$^{208}$Pb data, using a non-relativistic CHF method with a modified density
dependence of the functional. Therefore, it should be noted that K$_\infty$
cannot be extracted from the measurement of the GMR in a single nucleus:
several parts of the functionnal are tested simultaneously, namely
K$_\infty$, K$_{sym}$ and its density dependence.

A previous study on the role of superfluidity on nuclear incompressibility
has been performed, finding a negligible effect \cite{civ91}, but the
theoretical approach was not self-consistent. Indeed self-consistency is
crucial since pairing effects are expected to be small in the GMR: this high
energy mode is mainly built from particle-hole configurations located far
from the Fermi level, where pairing do not play a major role. However giant
resonances are known to be very collective \cite{har01} and pairing can
still have a sizable effect on the GMR properties: around 10 \% on the
centroid, which is the level of accuracy of present analysis on the
extraction of K$_\infty$ \cite{col04,vre03}. This requires the advent of
accurate microscopic models in the pairing channel, such as fully
self-consistent Quasiparticle Random Phase Approximation (QRPA)
\cite{jli08,paa03}, achieved only recently. Experimentally, the measurement
of the GMR on an isotopic chain facilitates the study of superfluidity on
the GMR properties \cite{tli07}, and the possibility to measure the GMR in
unstable nuclei emphasizes this feature \cite{mon08}.

It is therefore necessary to go towards the measurement of the GMR on
several nuclei, such as an isotopic chain. The overused method of precise
GMR measurements in a single nucleus, such as $^{208}$Pb, may not be the
relevant approach. Other nuclei have been used such as $^{90}$Zr and
$^{144}$Sm. Indeed when considering the available GMR data from which the
K$_\infty$ value has been extracted, $^{208}$Pb is stiffer than the Sn, Zr
and Sm nuclei: K$_\infty$ is about 20 MeV larger, both in non-relativistic
and in relativistic approaches \cite{col04,vre03}. The question may not be
"why Tin are so soft ?"\cite{pie07} but rather "why $^{208}$Pb is so stiff
?".

Recently the GMR was measured on the stable Tin istopic chain (from
$^{112}$Sn to $^{124}$Sn) \cite{tli07}. Once again it has been noticed that
it is not possible to describe the GMR both in Sn and in Pb with the same
functional, Tin beeing softer than Pb \cite{pie07,jli08}. In the non
relativistic case, fully self-consistent QRPA calculations on Sn isotopes
lead to K$_\infty$ $\simeq$ 215 MeV. The relativistic DDME2 paremeterisation
using QRPA describes well the GMR in the Sn isotopes \cite{lal09}, but
predict a low value of the GMR in $^{208}$Pb compared to the experiment, as
can be seen on Fig. 8 of Ref. \cite{vre03}. On the contrary, a recent
relativistic functional describes well the $^{208}$Pb GMR, but
systematically overestimates the Sn GMR values of about 1 MeV \cite{nik08}.
Finally, attempts have been performed in order to describe Sn GMR data with
relativistic fuctionals having a lower incompressibility and hence different
density dependence and K$_{sym}$ value \cite{pie09}. Once again the
$^{208}$Pb and Tin GMR cannot be described at the same time: this puzzling
situation is due to the higher value of K$_\infty$ extracted from
$^{208}$Pb, compared to Tin, Sm and Zr nuclei.

In Ref. \cite{jli08} it has been found that including pairing effects in the
description of the GMR allows to explain part of the Sn softness : pairing
may decrease the predicted centroid of the GMR of few hundreds of keV,
located at $\sim$ 16 MeV. This is sufficient to change by about 10 MeV the
extracted value of the incompressibility of nuclei K$_A$, defined as
\cite{bla80}:

\begin{equation}\label{eq:ka}
E_{GMR}=\sqrt{\frac{\hbar^2K_A}{m\langle r^2\rangle}}
\end{equation}
where m is the nucleon mass and $\langle$r$^2$$\rangle$ is the ground-state
mean square radius.

In this work we follow up this idea and show that the consequences of 
superfluidity on nuclear incompressibility may solve the above mentioned
puzzle. It should be noted that pairing is vanishing in the doubly magic
$^{208}$Pb nucleus, unlikely the other nuclei. It is necessary to use a
fully microsocopic method including an accurate pairing approach. In order
to predict the GMR in a microscopic way we use the constrained HF method,
extended to the full Bogoliubov pairing treatment (CHFB). The CHF(B) method
has the advantage to very precisely predict the centroid of the GMR using
the m$_{-1}$ sumrule \cite{boh79,col04b}. The whole residual interaction
(including spin-orbit and Coulomb terms) is taken into account and this
method is by construction the best to predict the GMR centroid
\cite{col04b}. Introducing the monopole operator as a constraint, the
m$_{-1}$ value is obtained from the derivative of the mean value of this
operator. The m$_1$ sumrule is extracted from the usual double commutator,
using the Thouless theorem \cite{tho61}. Finally the GMR centroid is given
by E$_{GMR}$=$\sqrt{m_1/m_{-1}}$. All details on the CHF method can be found
in \cite{boh79,col04}.

The extension of the CH method to the CHFB case uses the HFB approach in
coordinate space \cite{dob84}. Skyrme functionals and a zero-range surface
pairing interaction are used in this work. The magnitude of the pairing
interaction is adjusted so to describe the trend of the neutron pairing gap
in Tin isotopes. This interaction is known to describe a large variety of
pairing effects in nuclei \cite{ben03}. 

Fig. 1 displays the GMR energy obtained from the Sn measurements (times
A$^{1/3}$ to correct for the slow lowering of the GMR with the nuclear mass
\cite{har01}). Microscopic CHFB predictions using two functionals are also
shown: SLy4 \cite{cha98} (K$_\infty$=230 MeV, which describes well the Pb
GMR data), and SkM* \cite{bar82} (K$_\infty$=215 MeV). Without pairing, the
SLy4 interaction overestimates the Sn GMR data. Pairing effects (CHFB
calculations) decrease the centroid of the GMR, getting closer to the data.
This confirms the results of \cite{jli08}, where a self-consistent HFB+QRPA
approach was used to describe the Tin data. It should be noted that using a
less general BCS approach for the pairing channel increases the GMR energy
\cite{jli08,tse09,civ91}. This is due to the problematic treatment of high
energy single-particle states in the BCS model. It is therefore important to
use the full HFB approach to correctly describe pairing effects on the GMR,
especially for the pairing residual interaction. Fig. 1 shows, as expected,
that SkM* predictions with pairing effects are in better agreement with the Sn
data.

\begin{figure}[htb]
\includegraphics[width=8.5cm]{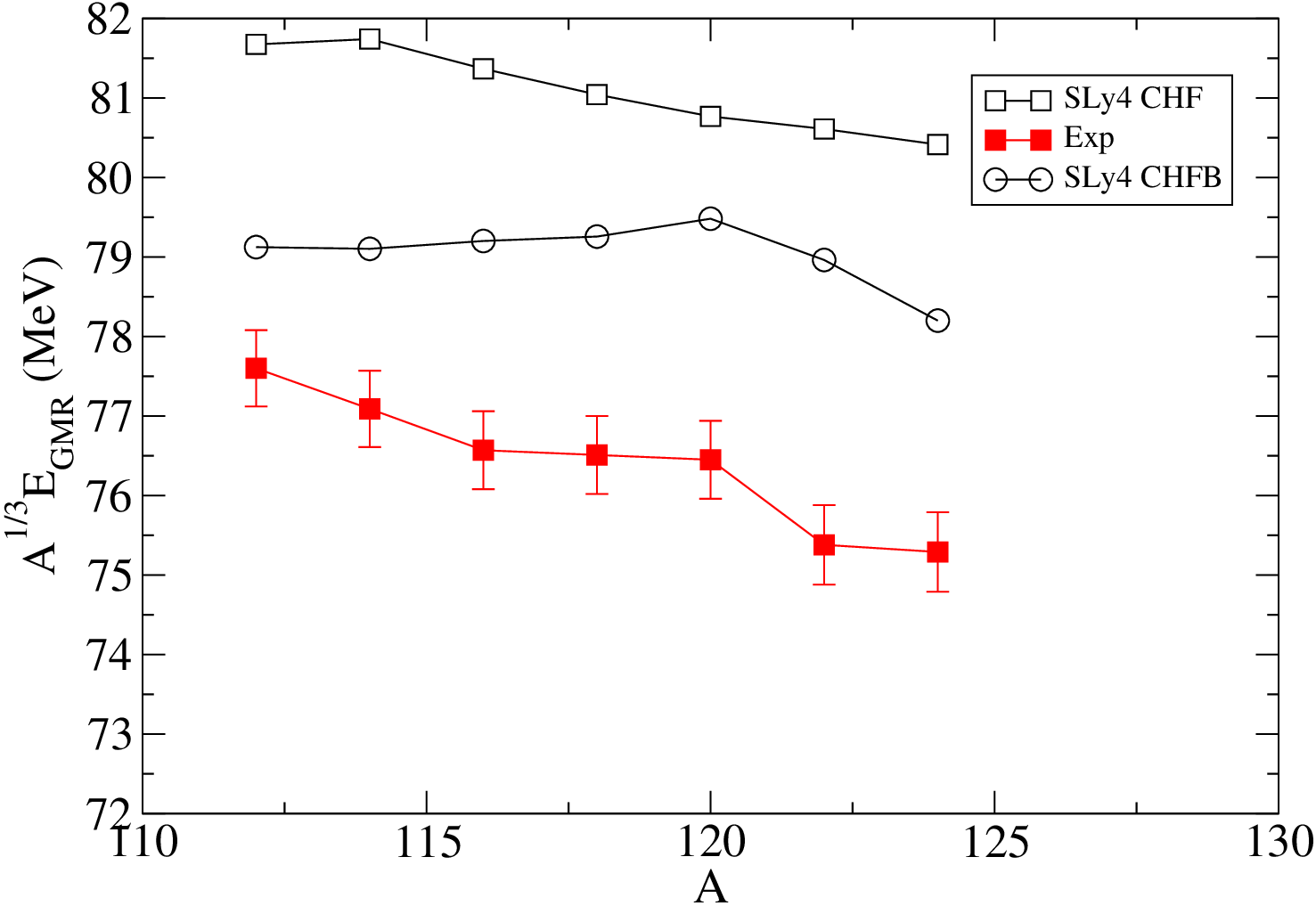}
\\
\includegraphics[width=8.5cm]{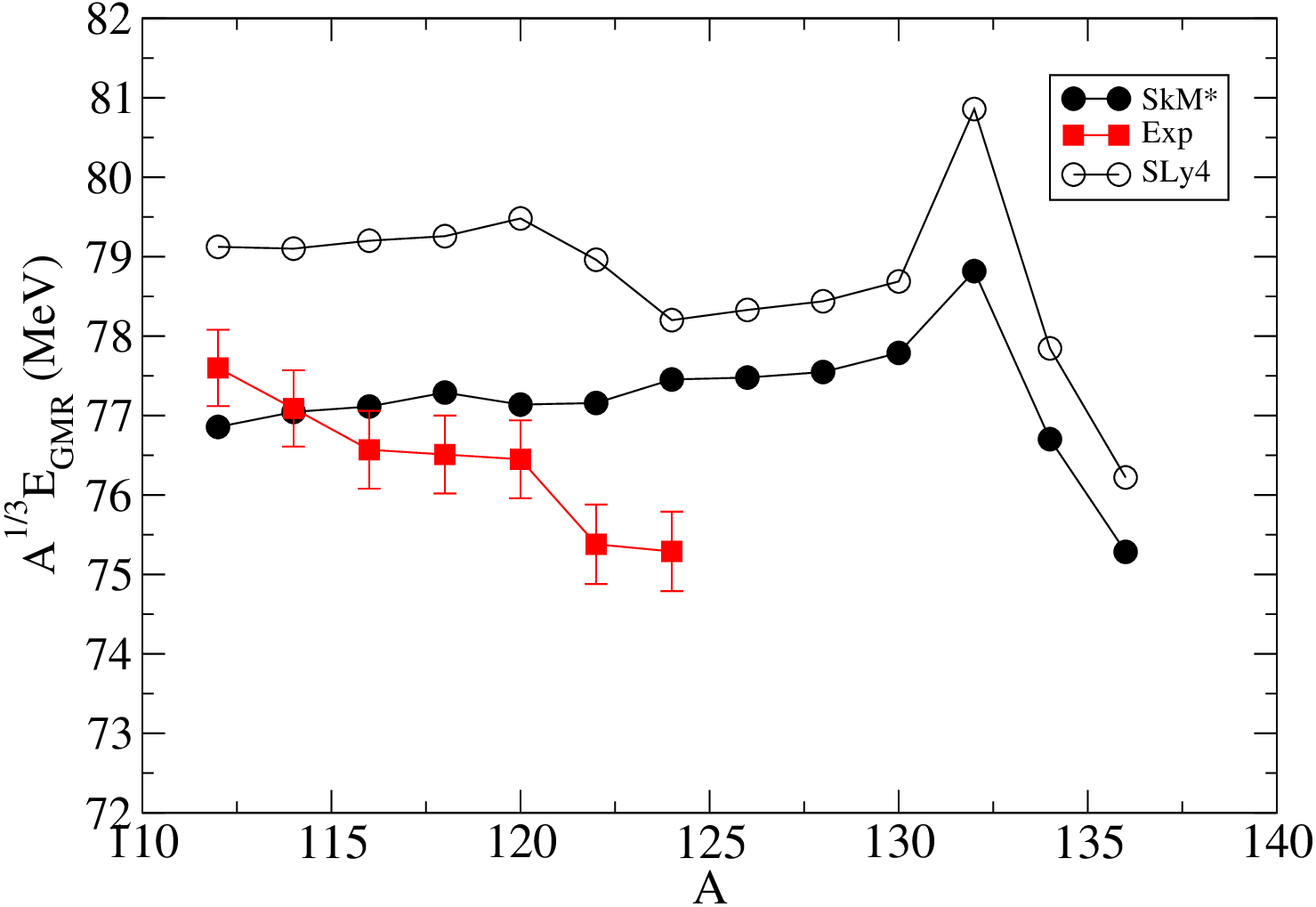}
\caption{{\it Top:} Excitations energies of the GMR in stable $^{112-124}$Sn isotopes
calculated with constrained HF and constrained HFB methods and the SLy4 interaction, 
compared to the data. {\it Bottom:} Same for $^{112-136}$Sn isotopes, using the
CHFB method and the SLy4 and SkM* interactions   
\label{fig:gmr}}
\end{figure}

The striking feature of Fig. 1 is the peak of the GMR centroid, located at
the doubly magic $^{132}$Sn nucleus, using the CHFB predictions. This
indicates that pairing effects should be considered to describe the
behaviour of nuclear incompressibility, and that vanishing of pairing make
the nuclei stiffer to compress, confirming our previous statement on the
stiffness of $^{208}$Pb. The importance of pairing effect can be understood
in a simple way: since the nuclear incompressibility is defined as the
second derivative of the energy functional at saturation density
\cite{bla80}, there is no obvious reason why the pairing terms of the
functional would play no role in the nuclear incompressibility. Nuclear
incompressibility is indeed very close from a residual interaction (as a
second derivative of the energy functionals with respect to the density),
and it is known that pairing effects are relevant in residual interactions
\cite{ben03}. This is straightforward in nuclear matter where K$_\infty$ is
expressed from the F$_0$ Landau parameter \cite{gle88}. However, on Fig. 1,
the GMR centroid is shifted to lower energies for more neutron-rich nuclei
than $^{132}$Sn, also because of the appearance of a soft L=0 mode,
predicted in QRPA calculations beyond $^{132}$Sn \cite{ter07}. 

To further investigate the role of pairing on nuclear incompressibility,
Fig. 2 displays K$_A$ (defined by Eq. (1)) with respect to the average
pairing gap calculated using the HFB approach, from $^{112}$Sn to
$^{132}$Sn. A clear correlation is observed: the more superfluid the nuclei,
the lower the incompressibility. Hence it may be easier to compress
superfluid nuclei. This may be the first evidence of the role of
superfluidity on the compressibility of a Fermionic system. A possible
interpretation is that Cooper pairs can modify bulk properties, as known
from nuclear physics phenomenology \cite{boh69}.

\begin{figure}[htb]
\includegraphics[width=8.5cm]{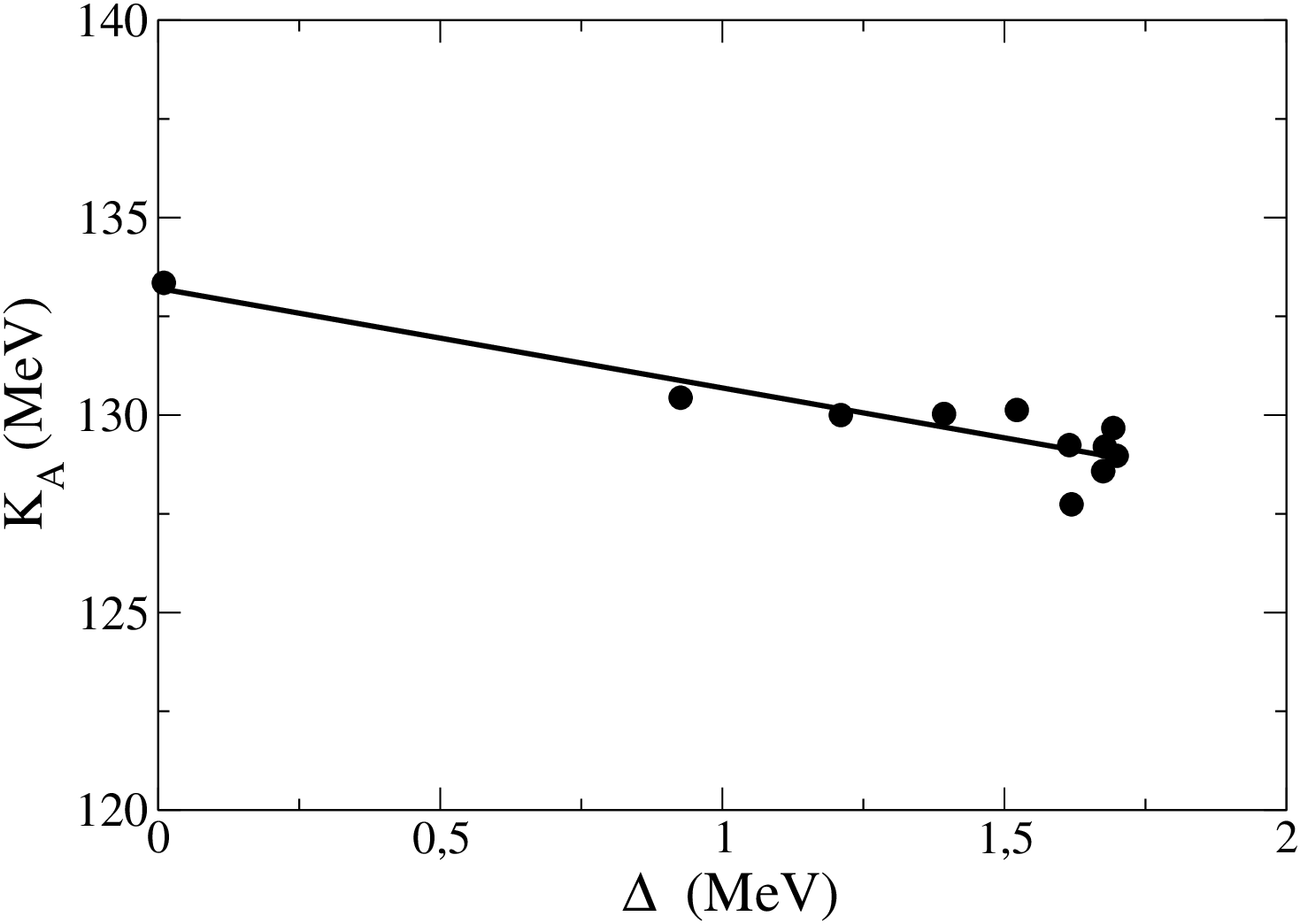}
\caption{Nuclear incompressibilities K$_A$ in $^{112-132}$Sn isotopes calculated
with the CHFB method and the SkM* interaction, as a function of the pairing
gap $\Delta$ predicted by the HFB calculation.  
\label{fig:ka}}
\end{figure}

The decrease of incompressibility in superfluid nuclei raises the question
of a similar effect in infinite nuclear matter: for now, incompressibility
is given independently from the pairing part of the functional. However,
considering present results, equations of state used for neutron star and
supernovae predictions should take into account pairing to provide their
incompressibility value. The comparison with GMR data shows, as mentioned
above, that the functional as a whole (including pairing effects) is probed.
The question of the behaviour of K$_\infty$ with respect to the pairing gap
is raised: it seems clear from nuclear data that nuclei incompressibility
K$_A$ decreases with increasing pairing gap. This should be investigated in
nuclear matter.

It should be noted that recent attempts have been performed to extract the
K$_{sym}$ value, and its corresponding quantity in nuclei (K$_\tau$): the
incompressibility in the Sn isotopic chain has been studied using the
macroscopic (liquid drop) formula of nuclei incompressibility K$_A$ derived
by Blaizot \cite{tli07,sag07,bla80}. However the effect of pairing
demonstrated above shows that the current macroscopic approach may not be
well designed: on Fig. 2, pairing effects induce $\sim$ 10 MeV change on the
nuclear incompressibility K$_A$. Hence the macroscopic expression of nuclei
incompressibility K$_A$ should be extended to these terms, since the
appropriate definition of nuclei incompressibility is the second derivative
of the microscopic energy functional \cite{bla80}. Presently, the
microscopic approach is more relevant to extract K$_{sym}$: the extraction
of K$_\infty$, K$_{sym}$ and the density dependence of the functional are
related, as stated in Ref. \cite{col04b}. The GMR data on isotopic chains
should be used on this purpose.

It is not possible to describe the GMR centroid of both $^{208}$Pb and other
nuclei with the same functional, as stated above. The puzzle of the
stiffness of $^{208}$Pb may come from its doubly magic behaviour. In Fig. 1
there is a sharp peak at doubly magic $^{132}$Sn, and it would be very
interesting to measure the GMR in this unstable nucleus. It should be noted
that such experiments are now feasible \cite{mon08}. A possible explanation
of the $^{208}$Pb stiffness is that the experimental data of E$_{GMR}$ is
especially increased in the case of doubly magic nuclei, as observed in
$^{208}$Pb compared to the GMR data available in other nuclei (such as the
Tin isotopic chain). This difficulty to describe with a single functional
both doubly magic and other nuclei has already been observed on the masses,
namely the so-called "mutually enhancement magicity" (MEM), described in
\cite{zel83,lun03}: functionals designed to describe masses of open-shell
nuclei cannot predict the masses of doubly magic nuclei such as $^{132}$Sn
and $^{208}$Pb, which are systematically more bound that predicted. In order
to consider MEM, it may be necessary to take into account quadrupole
correlation effects due to the flatness of the potentials for open-shell
nuclei \cite{ben05}. K$_A$ being related to the second derivative of the
energy with respect to the density, it would be useful to find a way to
predict the GMR beyond QRPA by taking into account such quadrupole
correlations. This may solve the current puzzle of the stiffness of
$^{208}$Pb. It should be noted that it would also be relevant to measure the
GMR on the Pb isotopic chain in order to provide a similar analysis than the
one on the Sn nuclei.

In conclusion, it is shown that superfluidity favours the compressibility of
nuclei, using a fully microscopic CHFB approach on the Tin isotopic chain.
This may be the first evidence of a sizable effect of superfludity on the
compressibility of a Fermionic system. Pairing effects should be described
using a full microscopic HFB treatment. Doubly magic nuclei exhibit a
specific increase of the GMR energy, due to the collapse of pairing.
$^{208}$Pb is therefore the "anomalous" data compared to the others. It is
not possible to disentangle pairing interaction from the equation of state
when providing the nuclear matter value of K$_\infty$. Indeed the pairing
gap dependence on the nuclear matter incompressibility should be
investigated, since it is shown that incompressibility decreases with
increasing pairing gap in nuclei. Additional theoretical investigations are
called for in order to predict the GMR including the mutually enhancement
magicity effect. The macroscopic extraction of K$_{sym}$ may be ill-defined
and should be extended to include pairing effects. Experimentally,
measurements of the GMR in unstable nuclei should be performed in doubly
magic $^{132}$Sn, as well as extending the measurement on the Sn and Pb
isotopic chains.

The author thanks G. Col\`o, M. Grasso, J. Margueron, Nguyen Van Giai, P.
Ring, H. Sagawa and M. Urban for fruitful discussions about the results of
this work.

\end{document}